\documentclass[superscriptaddress,groupedaddress,nofootinbib]{revtex4}

\usepackage{graphicx}  
\usepackage{dcolumn}   
\usepackage{bm}        
\usepackage{amssymb}   
\usepackage{color}
\begin{document}

\title{Scalar--multi-tensorial equivalence for higher order $f\left(  R,\nabla_{\mu}%
R,\nabla_{\mu_{1}}\nabla_{\mu_{2}}R,...,\nabla_{\mu_{1}}...\nabla_{\mu_{n}%
}R\right)  $ theories of gravity}

\author{R. R. Cuzinatto}
\email{rodrigo.cuzinatto@unifal-mg.edu.br}
\affiliation{Instituto de Ci\^{e}ncia e Tecnologia, Universidade Federal de Alfenas.\\Rod. Jos\'{e} Aur\'{e}lio Vilela (BR 267), Km 533, n. 11999, CEP 37701-970, \\Po\c{c}os de Caldas, MG, Brazil. }
\author{C. A. M. de Melo}
\email{cassius.melo@unifal-mg.edu.br}
\affiliation{Instituto de Ci\^{e}ncia e Tecnologia, Universidade Federal de Alfenas.\\Rod. Jos\'{e} Aur\'{e}lio Vilela (BR 267), Km 533, n. 11999, CEP 37701-970, \\Po\c{c}os de Caldas, MG, Brazil. }
\affiliation{Instituto de F\'{\i}sica Te\'{o}rica, Universidade Estadual Paulista.\\Rua Bento Teobaldo Ferraz 271 Bloco II, P.O. Box 70532-2, CEP 01156-970, \\S\~{a}o Paulo, SP, Brazil.}
\author{L. G. Medeiros}
\email{leogmedeiros@ect.ufrn.br}
\affiliation{Instituto de F\'{\i}sica Te\'{o}rica, Universidade Estadual Paulista.\\Rua Bento Teobaldo Ferraz 271 Bloco II, P.O. Box 70532-2, CEP 01156-970, \\S\~{a}o Paulo, SP, Brazil.}
\affiliation{Escola de Ci\^{e}ncia e Tecnologia, Universidade Federal do Rio Grande do Norte.\\Campus Universit\'{a}rio, s/n - Lagoa Nova, CEP 59078-970, \\Natal, RN, Brazil. }

\author{P. J. Pompeia}
\email{pompeia@ita.br}
\affiliation{Departamento de F\'{\i}sica, Instituto Tecnol\'{o}gico de Aeron\'{a}utica.\\Pra\c{c}a Mal. Eduardo Gomes 50, CEP 12228-900, \\S\~{a}o Jos\'{e} dos Campos, SP, Brazil. }


\begin{abstract}
The equivalence between theories depending on the derivatives of $R$, i.e.
$f\left(  R,\nabla R,...,\nabla^{n}R\right) $, and scalar--multi-tensorial
theories is verified. The analysis is done in both metric and Palatini formalisms.
It is shown that $f\left(  R,\nabla R,...,\nabla^{n}R\right)  $ theories
are equivalent to scalar--multi-tensorial ones resembling Brans-Dicke theories with kinetic terms $\omega_{0}=0$ and $\omega_{0}= - \frac{3}{2}$ for metric and Palatini formalisms respectively. This result is analogous to what happens for $f(R)$ theories.  It is worthy emphasizing that the scalar--multi-tensorial theories obtained here differ from Brans-Dicke ones due to the presence of multiple tensorial fields absent in the last. Furthermore, sufficient conditions are established for $f\left(  R,\nabla R,...,\nabla^{n}R\right)  $ theories to be written as scalar--multi-tensorial theories. Finally, some examples are studied and the comparison of $f\left(  R,\nabla R,...,\nabla^{n}R\right)  $ theories to $f\left(  R,\Box R,...\Box^{n}R\right)  $ theories is performed.
\end{abstract}

\maketitle

\section{Introduction}

Alternative scenarios for the standard theory of gravitation have long been proposed in order to circumvent several problems presented by General Relativity (renormalization \cite{Utiyama1962,Stelle1977}, inflation \cite{Guth1981,Linde1990}, the present day accelerated expansion of the universe \cite{Astier2006,Riess2004}, and so on). As examples of these proposals, one can cite
quadratic Lagrangians on the Riemann tensor theories
\cite{Querella1998,Buch1987}, the Horava--{}Lifshitz model \cite{Horava2009},
braneworld models \cite{Randall1999,Sundrum1999,Papan2002,Cuzinatto2014},
$\Lambda$CDM model \cite{Carroll2001,Padmanabhan2003}, etc..

Among these propositions, three are of special interest here. The first one is $f\left(  R\right)  $ theories
\cite{Faraoni2010,Felice2010,Odint2011,Capo2002,Capo2011,Gunther2005,Carroll2004,Odint2003}, which
were presented as an alternative scenario for the standard $\Lambda$CDM model
in an attempt to cure the cosmological constant problem \cite{Faraoni2010}. $f(R)$ gravity was also motivated as an alternative to dark energy models.

The second class of models that we are specially concerned with in this work is the scalar-tensor
theories \cite{BransDicke1961,Faraoni2004,Cassius,Cassius2}. In these theories, part of
the gravitational interaction is described by a scalar field. In the original
paper by Brans and Dicke \cite{BransDicke1961}, a scalar field was introduced along with the metric tensor in an attempt to implement Mach's principle. The interest in Brans-Dicke work was renewed with
string theory \cite{Capo2011}.

An important feature of both $f\left(  R\right)  $ and scalar-tensor theories
lies on the fact that they can be proved to be equivalent (at least at classical level),
i.e. a $f\left(R\right)  $ general model can be cast into the form of a Brans-Dicke theory with a potential
\cite{Barrow1988,Sotirou2006}.

The third category of interest here is the one of theories including
derivatives of the scalar curvature $R$ \cite{Odint2007,Odint2008,EPJC2008,Biswas2010,ASS2011,GRG2015,Biswas2015}.
They were inspired by string theory, or motivated by quantum loop corrections, or
as alternatives to dark energy models. From the point of view of quantum gravity,
terms containing derivatives of scalar curvature (and Riemann and Ricci tensors)
are necessary for the renormalizability of the theory \cite{Buchbinder1992,Asorey1997,Modesto2012,PRL2012},
although these terms usually produce unphysical massive ghost \cite{Shapiro2015,Modesto2016}.
Theories containing derivatives of $R$ can be seen either as toy models, or as effective theories, or even as full theories. In any case, the question addressed in this paper is: Is there a
scalar--multi-tensorial equivalent theory for this class of theories? As far as the
authors are aware, this question has been addressed for the particular case of
$f\left(  R,\Box R,...\Box^{n}R\right)  $ theories
\cite{Cecotti1987,Wands1994}. Here, the interest is devoted to a more general
category of theories, namely $f\left(  R,\nabla_{\mu}R,\nabla_{\mu_{1}}%
\nabla_{\mu_{2}}R,...,\nabla_{\mu_{1}}...\nabla_{\mu_{n}}R\right)  $ theories --- henceforth written as $f\left(  R,\nabla R,...,\nabla^{n}R\right)  $, or simply as $f (  R,\nabla R,...)$, for short-hand notation. As it shall be seen, besides the scalar field other auxiliary tensorial fields must be introduced. The analysis is restricted to a category of theories with
regular Hessian matrix (from this point of view, $f\left(  R,\Box
R,...\Box^{n}R\right)  $ theories are singular) and it is performed both in
the metric and Palatini formalisms.

With the introduction of the scalar--multi-tensorial structure, the problem of dealing with a single complicated higher-order field equation is substituted by the task of analyzing a larger number of field equations but with lower order of derivatives. This is known be useful in several situations. For instance, it is particularly efficient when one intends to perform numerical analyzes, as in \cite{GRG2015}. Moreover, it is simpler to examine the canonical structure of the theory and study its constraints when auxiliary fields are defined and a Lagrangian of lower order is considered \cite{Gitman1983,Bertin2008}. In addition, the procedure of order reduction may facilitate the scrutiny of the eventual unitarity character of the theory \cite{Kaparulin2014}.

The paper is organized as follows: In Section \textbf{2}, the equivalence of
$f\left(  R,\nabla R,...,\nabla^{n}R\right)  $ and scalar--multi-tensorial theories
is analyzed in the metric formalism. For the sake of clarity, the section starts with the analysis of the $f\left(  R,\nabla R\right)  $ case and is extended in the sequence. In
Section \textbf{3}, the analysis is repeated in the Palatini formalism. In
Section \textbf{4}, applications are performed and Section \textbf{5} is
devoted to our final remarks.

\section{Metric Formalism}

\subsection{Second order gravity theory}

Consider an action integral dependent on the scalar curvature and its first
derivative:
\begin{equation}
S=\int d^{4}x\sqrt{-g}\left[  f\left(  R,\nabla R\right)  +\mathcal{L}%
_{M}\right]  , \label{S 2ndOrder metric}%
\end{equation}
where $\mathcal{L}_{M}$ is the matter field Lagrangian.

Let $S^{\prime}$ be another action integral where a scalar and a vectorial
fields $\xi$ and $\xi_{\mu}$ are the fundamental fields and $R$ and its
derivative are considered as parameters:
\[
S^{\prime} = \int d^{4}x\sqrt{-g}\left[  f\left(  \xi,\xi_{\mu}\right)
-\frac{\partial f}{\partial\xi}\cdot\left(  \xi-R\right) \right. - \left. \frac{\partial
f}{\partial\xi_{\mu}}\cdot\left(  \xi_{\mu}-\nabla_{\mu}R\right)
+\mathcal{L}_{M}\right]  .
\]

The null variations of the action and independence of the variations in
$\xi,\,\xi_{\mu}$ lead to
\[
\left(
\begin{array}
[c]{cc}%
\frac{\partial^{2}f}{\partial\xi^{2}} & \frac{\partial^{2}f}{\partial
\xi\partial\xi_{\mu}}\\
\frac{\partial^{2}f}{\partial\xi_{\nu}\partial\xi} & \frac{\partial^{2}%
f}{\partial\xi_{\nu}\partial\xi_{\mu}}%
\end{array}
\right)  \left(
\begin{array}
[c]{c}%
\left(  \xi-R\right) \\
\left(  \xi_{\mu}-\nabla_{\mu}R\right)
\end{array}
\right)  =\left(
\begin{array}
[c]{c}%
0\\
0
\end{array}
\right)
\]
The new action $S^{\prime}$ will be equivalent to $S$ if
\begin{equation}
\det\left(
\begin{array}
[c]{cc}%
\frac{\partial^{2}f}{\partial\xi^{2}} & \frac{\partial^{2}f}{\partial
\xi\partial\xi_{\mu}}\\
\frac{\partial^{2}f}{\partial\xi_{\nu}\partial\xi} & \frac{\partial^{2}%
f}{\partial\xi_{\nu}\partial\xi_{\mu}}%
\end{array}
\right)  \neq0. \label{det2x2}%
\end{equation}
This condition leads to the following field equations
\begin{equation}
\left(
\begin{array}
[c]{c}%
\left(  \xi-R\right) \\
\left(  \xi_{\mu}-\nabla_{\mu}R\right)
\end{array}
\right)  =\left(
\begin{array}
[c]{c}%
0\\
0
\end{array}
\right)  , \label{equiv fields}%
\end{equation}
showing the equivalence of $S$ and $S^{\prime}$ under field equations. In
fact, from (\ref{equiv fields}): $\xi=R$\ and $\xi_{\mu}=\nabla_{\mu}R$.

It also becomes clear that $\frac{\partial f}{\partial\xi}$ and $\frac
{\partial f}{\partial\xi_{\mu}}$ are Lagrange multipliers in $S^{\prime}$.
This way, these quantities will be replaced by scalar and vectorial
fields, respectively,
\begin{eqnarray}
\phi &  \equiv\frac{\partial f}{\partial\xi},\label{phi}\\
\phi^{\mu}  &  \equiv\frac{\partial f}{\partial\xi_{\mu}}. \label{phi mu}%
\end{eqnarray}
Condition Eq.(\ref{det2x2}) ensures that $\xi=\xi\left(  \phi,\phi^{\mu
}\right)  $ and $\xi^{\nu}=\xi^{\nu}\left(  \phi,\phi^{\mu}\right)  $ exist.

With these quantities $S^{\prime}$ becomes
\[
S^{\prime} = \int d^{4}x\sqrt{-g}\left[  -\phi\xi-\phi^{\mu}\xi_{\mu}+f\left(
\xi,\xi_{\mu}\right) \right. + \left. \phi R+\phi^{\mu}\nabla_{\mu}R+\mathcal{L}_{M}\right].
\]
A potential $U\left(  \phi,\phi_{\mu}\right)  $ is defined as
\[
U\left(  \phi,\phi_{\mu}\right)  \equiv  \phi\xi\left(  \phi,\phi^{\mu}\right)
+\phi^{\mu}\xi_{\mu}\left(  \phi,\phi^{\mu}\right) -f\left(  \xi\left(
\phi,\phi^{\mu}\right)  ,\xi_{\mu}\left(  \phi,\phi^{\mu}\right)  \right)  .
\]

The action then reads:
\begin{equation}
S^{\prime}=\int d^{4}x\sqrt{-g}\left[  \phi R+\phi^{\mu}\nabla_{\mu}R-U\left(
\phi,\phi_{\mu}\right)  +\mathcal{L}_{M}\right]  , \label{Sprime(R,DR) metric}%
\end{equation}
which is the scalar-vectorial-tensorial equivalent theory to $S$. It is clear that no kinetic terms for $\phi$ and $\phi_{\mu}$ are present in $S^{\prime}$.

The coupling with the gradient of the Ricci scalar can be eliminated by using the identity
\[
\sqrt{-g}\phi^{\mu}\nabla_{\mu}R=\partial_{\mu}\left(  \sqrt{-g}\phi^{\mu
}R\right)  -\sqrt{-g}\nabla_{\mu}\phi^{\mu}R \, ,
\]
and expressing $S^{\prime}$, up to a surface term, as
\begin{equation}
S^{\prime}=\int d^{4}x\sqrt{-g}\left[  \Phi R-U\left(  \Phi,\phi_{\mu}%
,\nabla_{\mu}\phi^{\mu}\right)  +\mathcal{L}_{M}\right]  , \label{Sprime(Phi)}%
\end{equation}
where we have defined a new scalar field%

\begin{equation}
\Phi\equiv\left(  \phi-\nabla_{\mu}\phi^{\mu}\right)  , \label{Phi 2ndOrder}%
\end{equation}
and the potential%
\begin{equation}
U\left(  \Phi,\phi_{\mu},\nabla_{\mu}\phi^{\mu}\right)  \equiv\left(
\Phi+\nabla_{\mu}\phi^{\mu}\right)  \xi+\phi^{\mu}\xi_{\mu}-f\left(  \xi
,\xi_{\mu}\right)  . \label{U 2ndOrder metric}%
\end{equation}
Notice that potential $U$ depends on the new scalar $\Phi$ and the vector field $\phi_{\mu}$.

The theory established by Eq. (\ref{Sprime(Phi)}) resembles a Brans-Dicke theory,
\[
S_{BD}=\int d^{4}x\sqrt{-g}\left[  \varphi R-\frac{\omega_{0}}{\varphi}%
\nabla_{\mu}\varphi\nabla^{\mu}\varphi+\mathcal{L}_{M}\right]  ,
\]
with $\omega_{0}=0$ (which means that no explicit kinetic term for $\Phi$ is
present) and a potential for the scalar field, the vector field and its
covariant divergence.

\subsection{Extension to higher order gravity theories}

Consider the general action dependence on the scalar curvature and its
derivatives up to order $n$:
\[
S=\int d^{4}x\sqrt{-g}\left[  f\left(  R,\nabla R,\nabla^{2}R,...,\nabla
^{n}R\right)  +\mathcal{L}_{M}\right]  .
\]

A new action $S^{\prime}$ is proposed where scalar and tensorial fields are
introduced, replacing $R$ and its derivatives. In order to recast the original
theory in this new formulation, Lagrange multipliers are introduced so that
the new action is
\begin{eqnarray*}
 S^{\prime} & = &   \int d^{4}x\sqrt{-g}\left[  f\left(  \xi,\xi_{\mu},\xi_{\mu
\nu},...,\xi_{\mu_{1}...\mu_{n}}\right)\right.
- \frac{\partial f}{\partial\xi}  \cdot\left(  \xi-R\right) -\frac{\partial f}{\partial\xi_{\mu}}\cdot\left(  \xi_{\mu}-\nabla_{\mu}R\right) \\
 & - & \left.    \frac{\partial f}{\partial\xi_{\mu\nu}}\cdot\left(
\xi_{\mu\nu}-\nabla_{\mu}\nabla_{\nu}R\right) \right.
 - \left. ...-\frac{\partial f}%
{\partial\xi_{\mu_{1}...\mu_{n}}}\cdot\left(  \xi_{\mu_{1}...\mu_{n}}%
-\nabla_{\mu_{1}}...\nabla_{\mu_{n}}R\right)  +\mathcal{L}_{M}\right]  .
\end{eqnarray*}
The null variations of the action and the independence of the variations in
$\xi,\,\xi_{\mu},...,\xi_{\mu_{1}...\mu_{n}}$ lead to
\[
H\left(
\begin{array}
[c]{c}%
\left(  \xi-R\right) \\
\left(  \xi_{\mu}-\nabla_{\mu}R\right) \\
\left(  \xi_{\mu\nu}-\nabla_{\mu}\nabla_{\nu}R\right) \\
\vdots\\
\left(  \xi_{\mu_{1}...\mu_{n}}-\nabla_{\mu_{1}}...\nabla_{\mu_{n}}R\right)
\end{array}
\right)  =\left(
\begin{array}
[c]{c}%
0\\
0\\
0\\
\vdots\\
0
\end{array}
\right)
\]
where
\[
H\equiv\left(
\begin{array}
[c]{cccc}%
\frac{\partial^{2}f}{\partial\xi^{2}} & \frac{\partial^{2}f}{\partial
\xi\partial\xi_{\mu}} & \cdots & \frac{\partial^{2}f}{\partial\xi\partial\xi_{\mu_{1}...\mu_{n}}}\\
\frac{\partial^{2}f}{\partial\xi_{\nu}\partial\xi} & \frac{\partial^{2}%
f}{\partial\xi_{\nu}\partial\xi_{\mu}} &  \cdots & \frac{\partial^{2}f}{\partial\xi_{\nu}\partial\xi_{\mu_{1}...\mu_{n}}}\\
\vdots & \vdots  \cdots & \vdots\\
\frac{\partial^{2}f}{\partial\xi_{\nu_{1}...\nu_{n}}\partial\xi} &
\frac{\partial^{2}f}{\partial\xi_{\nu_{1}...\nu_{n}}\partial\xi_{\mu}} &
\cdots & \frac{\partial^{2}f}{\partial\xi_{\nu_{1}...\nu_{n}}\partial\xi
_{\mu_{1}...\mu_{n}}}%
\end{array}
\right)  .
\]
The new action $S^{\prime}$ will be equivalent to $S$ if the determinant of
the Hessian matrix $H$ is non-null, which leads to the following field equations
\[
\left(
\begin{array}
[c]{c}%
\left(  \xi-R\right) \\
\left(  \xi_{\mu}-\nabla_{\mu}R\right) \\
\left(  \xi_{\mu\nu}-\nabla_{\mu}\nabla_{\nu}R\right) \\
\vdots\\
\left(  \xi_{\mu_{1}...\mu_{n}}-\nabla_{\mu_{1}}...\nabla_{\mu_{n}}R\right)
\end{array}
\right)  =\left(
\begin{array}
[c]{c}%
0\\
0\\
0\\
\vdots\\
0
\end{array}
\right)  ,
\]
under which $S$ and $S^{\prime}$ become equivalent.

Now, let new tensorial quantities $\phi^{\left(  n\right)  }=\left\{
\phi,\phi^{\mu},...,\phi^{\mu_{1}...\mu_{n}}\right\} $ be introduced:
\begin{equation}
\left\{
\begin{array}
[c]{l}%
\phi\equiv\frac{\partial f}{\partial\xi},\\
\phi^{\mu}\equiv\frac{\partial f}{\partial\xi_{\mu}},\\
\phi^{\mu\nu}\equiv\frac{\partial f}{\partial\xi_{\mu\nu}},\\
\vdots\\
\phi^{\mu_{1}...\mu_{n}}\equiv\frac{\partial f}{\partial\xi_{\mu_{1}...\mu
_{n}}}.
\end{array}
\right.  \label{phi fields}%
\end{equation}

With these quantities $S^{\prime}$ becomes
\begin{eqnarray*}
S^{\prime}  &  = & \int d^{4}x\sqrt{-g}\left[  -\phi\xi-\phi^{\mu}\xi_{\mu}%
-\phi^{\mu\nu}\xi_{\mu\nu} \right. -  ...-\phi^{\mu_{1}...\mu_{n}}\xi_{\mu_{1}...\mu_{n}}
+f\left(  \xi,\xi_{\mu},\xi_{\mu\nu},...,\xi_{\mu_{1}...\mu_{n}}\right) \\
& + &  \phi R+\phi^{\mu}\nabla_{\mu}R+\phi^{\mu\nu}\nabla_{\mu}%
\nabla_{\nu}R + \left....+\phi^{\mu_{1}...\mu_{n}}\nabla_{\mu_{1}}...\nabla_{\mu_{n}%
}R+\mathcal{L}_{M}\right]  ,
\end{eqnarray*}
since the condition $\det H\neq0$ ensures that $\xi=\xi\left(  \phi,\phi^{\mu
},...\right)  ,...,\xi^{\mu_{1}...\mu_{n}}=\xi^{\mu_{1}...\mu_{n}}\left(
\phi,\phi^{\mu},...\right)  $ exist.

Define the potential
\begin{equation}
U\left( \phi, \phi_{\mu},\phi_{\mu\nu},...,\phi_{\mu_{1}...\mu_{n}}\right)
 = \phi\xi+\phi^{\mu}\xi_{\mu}+\phi^{\mu\nu}\xi_{\mu\nu}
 + ...+\phi^{\mu_{1}%
...\mu_{n}}\xi_{\mu_{1}...\mu_{n}} -f\left(  \xi,\xi_{\mu},\xi_{\mu\nu}%
,...,\xi_{\mu_{1}...\mu_{n}}\right). \label{U(phi fields)}%
\end{equation}
The action is then cast into the form
\[
S^{\prime}  = \int d^{4}x\sqrt{-g}\left[  -U\left(  \phi,\phi_{\mu},\phi
_{\mu\nu},...,\phi_{\mu_{1}...\mu_{n}}\right)  \right.  \\
 +  \phi R+\phi^{\mu}\nabla_{\mu}R+\phi^{\mu\nu}\nabla_{\mu}%
\nabla_{\nu}R - \left....+\phi^{\mu_{1}...\mu_{n}}\nabla_{\mu_{1}}...\nabla_{\mu_{n}%
}R+\mathcal{L}_{M}\right]  ,
\]
which is the scalar--multi-tensorial equivalent theory to $S$. The action $S^{\prime}$ above can be
rewritten by considering that
\begin{equation}
 \sqrt{-g}\phi^{\mu_{1}...\mu_{n}}\nabla_{\mu_{1}}...\nabla_{\mu_{n}}%
R= \sqrt{-g}\left(  -1\right)  ^{n}\nabla_{\mu_{n}}...\nabla_{\mu_{1}}\phi
^{\mu_{1}...\mu_{n}}R+\partial_{\mu}S_{\left(  n\right)  }^{\mu}%
.\label{TransfDeriv}%
\end{equation}
The last term of (\ref{TransfDeriv}) turns out to be a surface term when this
equation is substituted in the action integral. It
follows, up to surface terms,
\begin{equation}
S^{\prime} = \int d^{4}x\sqrt{-g}\left[  \Phi R \right.
- U\left(  \Phi,\phi_{\mu},...,\phi_{\mu_{1}...\mu_{n}},\nabla_{\mu}\phi^{\mu},...,\nabla_{\mu_{n}}...\nabla_{\mu_{1}}\phi^{\mu_{1}...\mu_{n}}\right)
+ \left. \mathcal{L}_{M}\right],\label{S prime gen metric}%
\end{equation}
where%
\begin{equation}
\Phi\equiv\phi-\nabla_{\mu}\phi^{\mu}+...+\left(  -1\right)  ^{n}\nabla
_{\mu_{n}}...\nabla_{\mu_{1}}\phi^{\mu_{1}...\mu_{n}},\label{Phi}%
\end{equation}
and%
\begin{eqnarray}
U &= &  U\left(  \Phi,\phi_{\mu},...,\phi_{\mu_{1}...\mu_{n}},\nabla_{\mu}%
\phi^{\mu},...,\nabla_{\mu_{n}}...\nabla_{\mu_{1}}\phi^{\mu_{1}...\mu_{n}%
}\right)  \nonumber\\
&= &  \left(  \Phi+\nabla_{\mu}\phi^{\mu}+...+\left(  -1\right)  ^{n+1}%
\nabla_{\mu_{n}}...\nabla_{\mu_{1}}\phi^{\mu_{1}...\mu_{n}}\right)
\xi\nonumber\\
&  + &\phi^{\mu}\xi_{\mu}+\phi^{\mu\nu}\xi_{\mu\nu}+...+\phi^{\mu_{1}...\mu_{n}%
}\xi_{\mu_{1}...\mu_{n}} -f\left(  \xi,\xi_{\mu},\xi_{\mu\nu},...,\xi_{\mu_{1}...\mu_{n}}\right)  .\label{U(Phi)}%
\end{eqnarray}
Eq. (\ref{S prime gen metric}) generalizes action (\ref{Sprime(Phi)}) by the addition of multiple tensorial fields. It also resembles a Brans-Dicke theory with $\omega_{0}=0$ and a potential depending on extra tensorial fields usually absent in the Brans-Dicke description. As
before, no kinetic term is present for $\Phi$ and in order to introduce it,
the Palatini formalism has to be considered. This is done in the next section. Meanwhile, the field equations are derived.

By varying the action with respect to the independent fields $g_{\mu \nu}$, $\Phi$, $\phi^{\mu}$,..., $\phi^{\mu_{1},...,\mu_{n}}$ and $\psi$, one obtains:
\begin{eqnarray}
&&\Phi G_{\mu\nu}-\left(  \nabla_{\mu}\nabla_{\nu}\Phi-g_{\mu\nu}\Box
\Phi\right)  +\frac{1}{2}g_{\mu\nu}U-\upsilon_{\mu\nu}=\kappa T_{\mu\nu} \,
,\label{eq mot dg metric}\\
&& R=\frac{\partial U}{\partial\Phi} \, ,\label{eq mot dPhi metric}\\
&& \frac{\partial U}{\partial\phi^{\rho}}-\nabla_{\mu}\frac{\partial
U}{\partial\left(  \nabla_{\mu}\phi^{\rho}\right)  }%
=0 \, ,\label{eq mot dphi vec metric}\\
&&\vdots  \nonumber\\
&&\frac{\partial U}{\partial\phi^{\rho_{1}...\rho_{n}}} +\left(  -1\right)
^{n}\nabla_{\mu_{n}}...\nabla_{\mu_{1}}\frac{\partial U}{\partial\left(
\nabla_{\mu_{1}}...\nabla_{\mu_{n}}\phi^{\rho_{1}...\rho_{n}}\right)
}=0 \,  ,\label{eq mot dphi tens metric}\\
&&\frac{\delta\mathcal{L}_{M}}{\delta\psi}=0 \, . \label{eq mot dpsi metric}%
\end{eqnarray}
where $\upsilon_{\mu\nu}\equiv\frac{\delta U}{\delta g^{\mu\nu}}$; $T_{\mu\nu}\equiv\frac{1}{2\kappa}\left[  g_{\mu\nu}\mathcal{L}_{M}\left(
g,\psi\right)  -2\frac{\delta\mathcal{L}_{M}}{\delta g^{\mu\nu}}\right]  $ is
the energy-momentum tensor obtained from the matter Lagrangian $\mathcal{L}%
_{M}$; $\psi$ is the matter field. General relativity is recovered when $\Phi
=1$, $\phi^{\rho}=\phi^{\rho_{1}\rho_{2}}=...=\phi^{\rho_{1}...\rho_{n}}=0$.
The set of equations above is the generalization of the
scalar-tensorial version of $f(R)$ theory in the metric formalism
\cite{Faraoni2010}.

The absence of a kinetic term for $\Phi$ does not imply that
this field carries no dynamics. The coupling of $\Phi$ with $R$ leads to field
equations where the dynamics for $\Phi$ become manifest: If the trace of Eq.~(\ref{eq mot dg metric}) is considered, then Eq.~(\ref{eq mot dPhi metric}) can be rewritten as
\[
3\Box\Phi+2U-\Phi\frac{\partial U}{\partial\Phi}-g^{\mu\nu}\upsilon_{\mu\nu} = \kappa T \, .
\]
A completely analogous result appears in metric $f(R)$ gravity \cite{Faraoni2010}.

\section{Palatini Formalism}

Now the Palatini formalism is developed. In this approach, the connection and
the metric are considered as independent fields. As a consequence, the
variations of the action will be taken with respect to both $\Gamma$ and
$g^{\mu\nu}$. Moreover, the matter fields Lagrangian $\mathcal{L}_{M}$ does
not explicitly depend on the connection. This is a necessary condition to
recover GR under the particular choice $f\left(  R,\nabla R\right)  =R$. This condition implies that all derivation operators present in the action
are built with the Levi-Civita connection.

Before proceeding to the general
Lagrangian $f\left(  R,\nabla R,\nabla^{2}R,...,\nabla^{n}R\right)  $, the
case $f\left(  R,\nabla R\right)  $ will be studied for clarification of
the steps to be followed when the most general case is analyzed.

\subsection{Second Order Theory}

\subsubsection{Field equations \label{Sec-Palatini-2ndOrder-FieldEq}}

The action for this theory is
\begin{equation}
S=\int d^{4}x\sqrt{-g}\left[  f\left(  \mathcal{R},\nabla\mathcal{R}\right)
+\mathcal{L}_{M}\left(  g,\psi\right)  \right]  , \label{S 2ndOrder Pal}%
\end{equation}
where $\nabla_{\rho} $ is the covariant derivative constructed with Christoffel
symbols $\left\{  _{\rho\sigma}^{\tau}\right\}  =\frac{1}{2}g^{\tau\lambda
}\left(  \partial_{\sigma}g_{\lambda\rho}+\partial_{\rho}g_{\sigma\lambda
}-\partial_{\lambda}g_{\rho\sigma}\right)  $. Quantity $\mathcal{R}$ is the
scalar curvature obtained from the general connection $\Gamma_{\rho\sigma
}^{\tau}$, i.e.
\begin{equation}
\mathcal{R}=g^{\mu\nu}\mathcal{R}_{\mu\nu}=g^{\mu\nu}\left(  \partial_{\rho
}\Gamma_{\mu\nu}^{\rho}-\partial_{\mu}\Gamma_{\rho\nu}^{\rho}+\Gamma_{\mu\nu
}^{\beta}\Gamma_{\rho\beta}^{\rho}-\Gamma_{\rho\nu}^{\beta}\Gamma_{\mu\beta
}^{\rho}\right)  . \label{R cal}%
\end{equation}
On the other hand, the Ricci scalar is $R=g^{\mu\nu}R_{\mu\nu}=g^{\mu\nu
}\left(  \partial_{\rho}\left\{  _{\mu\nu}^{\rho}\right\}  -\partial_{\mu
}\left\{  _{\rho\nu}^{\rho}\right\}  +\left\{  _{\mu\nu}^{\beta}\right\}
\left\{  _{\rho\beta}^{\rho}\right\}  -\left\{  _{\rho\nu}^{\beta}\right\}
\left\{  _{\mu\beta}^{\rho}\right\}  \right)  $. The variation of the action
integral is taken with respect to the metric tensor, the connection and the
matter field, leading to the following equations of motion%
\begin{eqnarray}
0  &  = &f^{\prime}\left(  \mathcal{R},\nabla\mathcal{R}\right)  \mathcal{R_{\mu
\nu}}-\frac{1}{2}f\left(  \mathcal{R},\nabla\mathcal{R}\right)  g_{\mu\nu
} + \left[  \frac{\delta\mathcal{L}_{M}}{\delta g^{\mu\nu}}-\frac{1}{2}g_{\mu
\nu}\mathcal{L}_{M}\left(  g,\psi\right)  \right] \, ,\label{eq mot 1}\\
0  &=& \bar{\nabla}_{\rho}\left(  \sqrt{-g}f^{\prime}\left(  \mathcal{R}%
,\nabla\mathcal{R}\right)  g^{\mu\nu}\right) \, ,\label{eq mot 2}\\
0  &=& \frac{\delta\mathcal{L}_{M}}{\delta\psi} \, , \label{eq mot 3}%
\end{eqnarray}
where we have defined
\begin{equation}
f^{\prime}\left(  \mathcal{R},\nabla\mathcal{R}\right)  \equiv\frac{\partial
f}{\partial\mathcal{R}}-\nabla_{\rho}\frac{\partial f}{\partial\nabla_{\rho
}\mathcal{R}} \, . \label{fprimesecond}%
\end{equation}
The bared covariant derivative $\bar{\nabla}$ is defined in terms of the
general connection: $\bar{\nabla}=\partial+\Gamma$.

The equation of motion resulting from the variation of the action with respect
to the connection is expressed in the form of Eq. (\ref{eq mot 2}) after we
use the identity
\begin{equation}
\bar{\nabla}_{\alpha}\left(  \sqrt{-g}f^{\prime}\left(  \mathcal{R}%
,\nabla\mathcal{R}\right)  g^{\alpha\mu}\right)  =0 \, , \label{trace of eq mot 2}%
\end{equation}
which is easily verified.

The conformal metric $h_{\mu\nu}$ is introduced:
\begin{equation}
h_{\mu\nu}\equiv f^{\prime}\left(  \mathcal{R},\nabla\mathcal{R}\right)
g_{\mu\nu} \, , \label{hmunu}%
\end{equation}
satisfying the following properties%
\[
\sqrt{-h}=\left[  f^{\prime}\left(  \mathcal{R},\nabla\mathcal{R}\right)
\right]  ^{2}\sqrt{-g},\quad\left(  h\equiv\det h_{\mu\nu}\right) \, ,
\]%
\[
h^{\alpha\beta}=\frac{1}{f^{\prime}\left(  \mathcal{R},\nabla\mathcal{R}%
\right)  }g^{\alpha\beta} \, ,
\]
and%
\begin{equation}
\bar{\nabla}_{\rho}\left(  \sqrt{-h}h^{\alpha\beta}\right)  =0\Rightarrow
\bar{\nabla}_{\rho}h_{\theta\lambda}=0 \, . \label{metricity}%
\end{equation}
The last expression is the metricity condition, which leads to the following
expression for the connection $\Gamma_{\;\mu\nu}^{\beta}$ (resembling the
Christoffel symbols with $g_{\mu\nu}$ replaced by $h_{\mu\nu}$):
\begin{equation}
\Gamma_{\;\mu\nu}^{\beta}=\frac{1}{2}h^{\alpha\beta}\left(  \partial_{\nu
}h_{\alpha\mu}+\partial_{\mu}h_{\nu\alpha}-\partial_{\alpha}h_{\mu\nu}\right)
. \label{Christ(h)}%
\end{equation}
With this expression, a relation between $\Gamma_{\;\mu\nu}^{\beta}$ and
$\left\{  _{\;\mu\nu}^{\beta}\right\}  $ can be established,
\begin{equation}
\Gamma_{\;\mu\nu}^{\beta}=\left\{  _{\;\mu\nu}^{\beta}\right\}  +\frac{1}%
{2}\frac{1}{f^{\prime}}g^{\alpha\beta}\left(  g_{\alpha\mu}\partial_{\nu
}f^{\prime}+g_{\nu\alpha}\partial_{\mu}f^{\prime}-g_{\mu\nu}\partial_{\alpha
}f^{\prime}\right)  , \label{connectionrelations}%
\end{equation}
Also a relation between $\mathcal{R}_{\mu\nu}$ and $R_{\mu\nu}$ is obtained:%
\begin{equation}
\mathcal{R}_{\mu\nu} = R_{\mu\nu}+\frac{3}{2}\frac{1}{\left(  f^{\prime}\right)
^{2}}\nabla_{\mu}f^{\prime}\nabla_{\nu}f^{\prime} \nonumber \\
- \frac{1}{2}\frac
{1}{f^{\prime}}\left(  \nabla_{\mu}\nabla_{\nu}f^{\prime}+\nabla_{\nu}%
\nabla_{\mu}f^{\prime}+g_{\mu\nu}\square f^{\prime}\right)  ,
\label{RicciIdentity}%
\end{equation}
where $\square=\nabla_{\rho}\nabla^{\rho}$.
For the scalar curvature:%
\begin{equation}
\mathcal{R}=R+\frac{3}{2}\frac{1}{\left(  f^{\prime}\right)  ^{2}}\left(
\nabla_{\mu}f^{\prime}\nabla^{\mu}f^{\prime}\right)  -3\frac{1}{f^{\prime}%
}\left(  \square f^{\prime}\right)  . \label{TraceRelation}%
\end{equation}

Eq. (\ref{RicciIdentity}) will be used in Eq. (\ref{eq mot 1}),
\begin{equation}
f^{\prime}\left(  \mathcal{R},\nabla\mathcal{R}\right)  \mathcal{R_{\mu\nu}%
}=\kappa T_{\mu\nu}+\frac{1}{2}g_{\mu\nu}f\left(  \mathcal{R},\nabla
\mathcal{R}\right)  \label{EqGrav}%
\end{equation}
where $T_{\mu\nu}$ is the energy-momentum tensor. Equivalently,
\begin{equation}
R_{\mu\nu}=\frac{\kappa}{f^{\prime}}T_{\mu\nu}+\frac{1}{2}g_{\mu\nu}\frac
{f}{f^{\prime}}-\frac{3}{2}\frac{1}{\left(  f^{\prime}\right)  ^{2}}%
\nabla_{\mu}f_{\nu}^{\prime}\nabla f^{\prime}
+\frac{1}{2}\frac{1}{f^{\prime}%
}\left(  \nabla_{\mu}\nabla_{\nu}f^{\prime}+\nabla_{\nu}\nabla_{\mu}f^{\prime
}+g_{\mu\nu}\square f^{\prime}\right)  \label{Rmunu}%
\end{equation}
or%
\begin{equation}
G_{\mu\nu}=\frac{\kappa}{f^{\prime}}T_{\mu\nu}-\frac{1}{2}g_{\mu\nu}\left(
\mathcal{R}-\frac{f}{f^{\prime}}\right)
 +\frac{1}{f^{\prime}}\left(
\nabla_{\mu}\nabla_{\nu}f^{\prime}-g_{\mu\nu}\square f^{\prime}\right)
-\frac{3}{2}\frac{1}{\left(  f^{\prime}\right)  ^{2}}\left(  \nabla_{\mu
}f^{\prime}\nabla_{\nu}f^{\prime}-\frac{1}{2}g_{\mu\nu}\nabla_{\beta}%
f^{\prime}\nabla^{\beta}f^{\prime}\right)  , \label{Gmunu}%
\end{equation}
where $G_{\mu\nu}\equiv R_{\mu\nu}-\frac{1}{2}g_{\mu\nu}R$. These are the
modified gravitational field equations. It reduces to the GR equations if
$f\left(  \mathcal{R},\nabla\mathcal{R}\right)  =R$.

\subsubsection{Scalar-Vectorial-Tensorial Theory}

We start with scalar and vector fields defined in the Palatini formalism:
\[
S=\int d^{4}x\sqrt{-g}\left[  f\left(  \mathcal{R},\nabla\mathcal{R}\right)
+\mathcal{L}_{M}\right]
\]

Proceeding exactly as in the metric approach (except that $\mathcal{R}$
appears instead of $R$) a new action $S^{\prime}$ is obtained as
\begin{equation}
S^{\prime}=\int d^{4}x\sqrt{-g}\left[  \Phi\mathcal{R}-U\left(  \Phi,\phi
_{\mu},\nabla_{\mu}\phi^{\mu}\right)  +\mathcal{L}_{M}\right]  ,
\label{Sprime(Phi) R cal}%
\end{equation}
where, $\Phi\equiv\left(  \phi-\nabla_{\mu}\phi^{\mu}\right)  $ and $U\left(
\Phi,\phi_{\mu},\nabla_{\mu}\phi^{\mu}\right)  \equiv\left(  \Phi+\nabla_{\mu
}\phi^{\mu}\right)  \xi+\phi^{\mu}\xi_{\mu}-f\left(  \xi,\xi_{\mu}\right)  $.
If $\mathcal{R}$ is replaced in terms of $R$, according to Eq.
(\ref{TraceRelation}), then we get, up to a surface term:
\begin{equation}
S^{\prime} = \int d^{4}x\sqrt{-g}\left[  \Phi R+\frac{3}{2}\frac{1}{\Phi}%
\nabla_{\mu}\Phi\nabla^{\mu}\Phi \right.
- \left. U\left(  \Phi,\phi_{\mu},\nabla_{\mu}%
\phi^{\mu}\right)  +\mathcal{L}_{M}\right] \, . \label{Sprime(Phi) Pal}%
\end{equation}
This is a theory that mimics a Brans-Dicke action with
$\omega_{0}=-\frac{3}{2}$. This result is quite similar to the one obtained for $f\left(  R\right)  $ theories {\cite{Faraoni2010}}, where $\frac{\partial
f}{\partial\mathcal{R}}$ is replaced by $f^{\prime}\left(  \mathcal{R}%
,\nabla\mathcal{R}\right)  $. As in the metric approach, an extra vector field is present.

\subsection{Generalization for higher derivatives}

Now the general case $f\left(  \mathcal{R},\nabla
_{\mu_{1}}\mathcal{R},...,\nabla_{\mu_{1}}...\nabla_{\mu_{n}}\mathcal{R}%
\right)  =f\left(  \mathcal{R},\nabla\mathcal{R},\nabla^{2}\mathcal{R}%
,...,\nabla^{n}\mathcal{R}\right)  $ will be analyzed.

\subsubsection{Field equations}

In this section, the previous results are generalized to Lagrangians depending
on higher derivatives of the curvature. Previously, it was checked that the
comparison of $f\left(  R,\nabla R\right)  $ theories with $f\left(
R\right)$ gravity led to the substitution $\frac{\partial f}{\partial\mathcal{R}%
}\rightarrow\frac{\partial f}{\partial\mathcal{R}}-\nabla_{\gamma}%
\frac{\partial f}{\partial\left(  \partial_{\gamma}\mathcal{R}\right)  }$. The
higher order Lagrangian is:
\begin{equation}
S = \int d^{4}x\sqrt{-g}\left[  f\left(  \mathcal{R}%
,\nabla_{\mu_{1}}\mathcal{R},...,\nabla_{\mu_{1}}...\nabla_{\mu_{n}%
}\mathcal{R}\right) \right.
 + \left. \mathcal{L}_{M}\left(  g,\psi\right)  \right] \, .
\label{S gen}%
\end{equation}
The abreviation $f\left(  \mathcal{R},\nabla\mathcal{R},\nabla^{2}%
\mathcal{R},...,\nabla^{n}\mathcal{R}\right)  =f\left(  \mathcal{R}%
,\nabla\mathcal{R},...\right)  $ will be used from now on, where there is no
risk of confusion.

Variations with respect to $g_{\mu\nu}$, $\Gamma$ and $\psi$ give results
completely analogous to the second order case: The equations of motion are
precisely Eqs. (\ref{eq mot 1}), (\ref{eq mot 2}) and (\ref{eq mot 3})
provided that we generalize $f^{\prime}\left(  \mathcal{R},\nabla
\mathcal{R}\right)  $\ to $f^{\prime}\left(  \mathcal{R},\nabla\mathcal{R}%
,...\right)  $\ as below:
\begin{equation}
f^{\prime}\left(  \mathcal{R},\nabla_{\rho_{1}%
}\mathcal{R},...,\nabla_{\rho_{1}}...\nabla_{\rho_{n}}\mathcal{R}\right)
=\frac{\partial f}{\partial\mathcal{R}} -\nabla_{\rho}\left(  \frac{\partial
f}{\partial\nabla_{\rho}\mathcal{R}}\right)  +...+\left(  -1\right)  ^{n}\nabla_{\rho_{n}}...\nabla_{\rho_{1}}%
\frac{\partial f}{\partial\nabla_{\rho_{1}}...\nabla_{\rho_{n}}\mathcal{R}} \, .
\label{fprimegeneral}%
\end{equation}
Relations Eq.(\ref{connectionrelations}), Eq.(\ref{RicciIdentity}) and
Eq.(\ref{TraceRelation}) and all results obtained
previously can be directly generalized just by taking
$f^{\prime}$ as the complete functional derivative --- Eq.(\ref{fprimegeneral}).
Now we turn to the problem of investigating the equivalence of the $f^{\prime
}\left(  \mathcal{R},\nabla\mathcal{R},...\right)  $-gravity theories with
scalar--multi-tensorial models.

\subsubsection{Scalar--multi-Tensorial Theory}

If the scalar, tensorial fields and the potential $U$ are defined as in the
metric approach -- Eqs. (\ref{phi fields}) and (\ref{U(phi fields)}) -- the
action integral (\ref{S gen}) takes the form
\begin{equation}
S^{\prime} = \int d^{4}x\sqrt{-g}\left[  \Phi\mathcal{R}\right. - U\left(  \Phi,\phi
_{\mu},...,\phi_{\mu_{1}...\mu_{n}},\nabla_{\mu}\phi^{\mu},...,\nabla_{\mu
_{n}}...\nabla_{\mu_{1}}\phi^{\mu_{1}...\mu_{2}}\right)
 + \left.\mathcal{L}_{M}\right] \, , \label{S prime gen Pal}%
\end{equation}
up to surface terms, where $\Phi$\ and $U\left(  \Phi,\phi_{\mu},...\right)  $
are those in Eqs. (\ref{Phi}) and (\ref{U(Phi)}). Eq. (\ref{S prime gen Pal})
describes a theory analogous to a Brans-Dicke theory with $\omega_{0}=-\frac{3}{2}$:
\[
S^{\prime}  = \int d^{4}x\sqrt{-g}\left[  \Phi R+\frac{3}{2}\frac{1}{\Phi
}\nabla_{\mu}\Phi\nabla^{\mu}\Phi\right.
- U\left(  \Phi,\phi_{\mu},...,\phi_{\mu_{1}...\mu_{n}},\nabla_{\mu
}\phi^{\mu},...,\nabla_{\mu_{n}}...\nabla_{\mu_{1}}\phi^{\mu_{1}...\mu_{n}%
}\right) + \left. \mathcal{L}_{M}\right] \, .
\]
Just like in the metric approach, extra tensorial fields are present establishing a significant difference with respect to Brans-Dicke theory.

The field equations are finally obtained for the scalar-tensor action:%
\begin{eqnarray}
&&  \Phi G_{\mu\nu}+\frac{3}{2}\frac{1}{\Phi}\left(  \nabla_{\mu}%
\Phi\nabla_{\nu}\Phi-\frac{1}{2}g_{\mu\nu}\nabla_{\rho}\Phi\nabla^{\rho}%
\Phi\right) -\left(  \nabla_{\mu}\nabla_{\nu}\Phi-g_{\mu\nu}\Box\Phi\right)
+\frac{1}{2}g_{\mu\nu}U-\upsilon_{\mu\nu}=\kappa T_{\mu\nu} \, , \label{eq mot dg Pal}\\
&&  \left.  R-3\frac{1}{\Phi}\square\Phi+\frac{3}{2}\frac{1}{\Phi^{2}}%
\nabla_{\rho}\Phi\nabla^{\rho}\Phi-\frac{\partial U}{\partial\Phi}=0\right. \, ,
\label{eq mot dPhi Pal}\\
&&  \left.  \frac{\partial U}{\partial\phi^{\rho}}-\nabla_{\mu}\frac{\partial
U}{\partial\left(  \nabla_{\mu}\phi^{\rho}\right)  }=0\right. \, ,
\label{eq mot dphi vec Pal}\\
&&  \vdots\nonumber\\
&&    \frac{\partial U}{\partial\phi^{\rho_{1}...\rho_{n}}}
 +\left(-1\right)  ^{n}\nabla_{\mu_{n}}...\nabla_{\mu_{1}}\frac{\partial U}%
{\partial\left(  \nabla_{\mu_{1}}...\nabla_{\mu_{n}}\phi^{\rho_{1}...\rho_{n}%
}\right)  }=0 \, , \label{eq mot dphi tens Pal}\\
&&  \left.  \frac{\delta\mathcal{L}_{M}}{\delta\psi}=0\right. \, ,
\label{eq mot dpsi Pal}%
\end{eqnarray}
where $\upsilon_{\mu\nu} = \frac{\delta U}{\delta g^{\mu\nu}}$.

As an opposition to the metric approach, the presence of the kinetic term for
$\Phi$ does not imply that this field carries dynamics. If one takes the trace of Eq.~(\ref{eq mot dg Pal}) and the resulting expression for $R$ is replaced on Eq.~(\ref{eq mot dPhi Pal}), one
finds
\[
2U-\Phi\frac{\partial U}{\partial\Phi}-g^{\mu\nu}\upsilon_{\mu\nu}=\kappa T \, ,
\]
It is clear that this is a constraint equation for $\Phi$. The same occurs in the context of $f\left(  R\right)  $ theories -- see e.g. Ref. \cite{Faraoni2010}.

\section{Application: the Starobinsky-Podolsky action}

The following system will be analyzed:
\begin{equation}
S=\int d^{4}x\sqrt{-g}\left[  R+\frac{c_{0}}{2}R^{2}+\frac{c_{1}}{2}%
\nabla_{\mu}R\nabla^{\mu}R+\mathcal{L}_{M}\right] \, , \label{action SP}%
\end{equation}
i.e.%
\begin{equation}
f\left(  R,\nabla_{\mu}R \right)  =R+\frac{c_{0}}%
{2}R^{2}+\frac{c_{1}}{2}\nabla_{\mu}R\nabla^{\mu}R \, , \label{f SP}%
\end{equation}
which implies
\[
f\left(  \xi,\xi_{\mu}\right)  =\xi+\frac{c_{0}}{2}\xi^{2}+\frac{c_{1}}{2}%
\xi_{\mu}\xi^{\mu}.
\]
The condition
\[
\det\left(
\begin{array}
[c]{cc}%
\frac{\partial^{2}f}{\partial\xi^{2}} & \frac{\partial^{2}f}{\partial
\xi\partial\xi_{\mu}}\\
\frac{\partial^{2}f}{\partial\xi_{\nu}\partial\xi} & \frac{\partial^{2}%
f}{\partial\xi_{\nu}\partial\xi_{\mu}}%
\end{array}
\right)  =\det\left(
\begin{array}
[c]{cc}%
c_{0} & 0\\
0 & c_{1}g^{\mu\nu}%
\end{array}
\right)  \neq0
\]
is satisfied as long as $c_{0}\neq0,\,c_{1}\neq0$. Under these constraints,
the theory from action (\ref{action SP}) is non-singular. Starobinsky-Podolsky
action could be made equivalent to a theory of the type $f\left(  R,\square
R\right)  $ up to a surface term after an integration by parts; however, the
resulting $f\left(  R,\square R\right)  $-theory would be singular.

\subsection{Metric formalism} \label{Sec-Metric-Sta-Pod}

The tensorial fields $\phi,\phi^{\mu}$ are
\begin{equation}
\left\{
\begin{array}
[c]{l}%
\phi\equiv\frac{\partial f}{\partial\xi}=\left(  1+c_{0}\xi\right)
\Rightarrow\xi=\frac{\left(  \phi-1\right)  }{c_{0}} \, ,\\
\phi^{\mu}\equiv\frac{\partial f}{\partial\xi_{\mu}}=c_{1}\xi^{\mu}%
\Rightarrow\xi^{\mu}=\frac{\phi^{\mu}}{c_{1}} \, ,
\end{array}
\right.  \label{fields SP}%
\end{equation}
and the potential $U$ is given by
\begin{equation}
U\left(  \phi,\phi_{\mu},\phi_{\mu\nu}\right)  =\phi\xi+\phi^{\mu}\xi_{\mu
}-f\left(  \xi,\xi_{\mu}\right)
 = \frac{\left(  \phi-1\right)  ^{2}}{2c_{0}%
}+\frac{\phi^{\mu}\phi_{\mu}}{2c_{1}} \, . \label{U SP}%
\end{equation}

The action integral $S^{\prime}=S^{\prime}\left(  \phi,\phi^{\mu}%
,R,\nabla_{\mu}R\right)  $ is promptly obtained by substituting Eq.
(\ref{U SP}) into (\ref{Sprime(R,DR) metric}). Then, one uses the definition
of $\Phi$, Eq. (\ref{Phi 2ndOrder}), to obtain:
\[
S^{\prime}=\int d^{4}x\sqrt{-g}\left[  \Phi R-\frac{1}{2c_{0}}\left(
\Phi+\nabla_{\mu}\phi^{\mu}-1\right)  ^{2} \right.
- \left. \frac{\phi^{\mu}\phi_{\mu}}{2c_{1}%
}+\mathcal{L}_{M}\right] \, .
\]

By extracting the variation of this action, one gets the field equations as
being precisely Eqs. (\ref{eq mot dg metric}), (\ref{eq mot dPhi metric}),
(\ref{eq mot dphi vec metric}) and (\ref{eq mot dpsi metric}) with Eq.
(\ref{U SP}) replacing $U$. Combining the equation of motion obtained in the
way described previously, results in the following set of coupled equations
for the scalar-vectorial part of $S^{\prime}$:
\[
\left\{
\begin{array}
[c]{l}%
-3\Box\Phi-\kappa T-2\left[  \frac{1}{2c_{0}}\left(  \Phi+\nabla_{\nu}%
\phi^{\nu}-1\right)  ^{2}+\frac{\phi^{\mu}\phi_{\mu}}{2c_{1}}\right]
+\Phi\frac{1}{c_{0}}\left(  \Phi+\nabla_{\beta}\phi^{\beta}-1\right)-\upsilon  =0 \, ,\\
\\
\frac{\phi_{\mu}}{c_{1}}-\nabla_{\mu}\left[  \frac{1}{c_{0}}\left(
\Phi+\nabla_{\gamma}\phi^{\gamma}-1\right)  \right]  =0 \, ,
\end{array}
\right.
\]
where
\begin{equation}
\upsilon=2\nabla_{\gamma}\left(\frac{1}{c_{0}}\left(\Phi+\nabla_{\sigma}\phi^{\sigma}-1\right)\phi^{\gamma}\right)+\frac{\phi^{\mu}\phi_{\mu}}{2c_{1}}.\label{vStarPod}
\end{equation}

The action (\ref{action SP}) can be rewritten, up to surface terms, as
\begin{equation}
S=\int d^{4}x\sqrt{-g}\left[  R+\frac{c_{0}}{2}R^{2}-\frac{c_{1}}{2}%
R\Box R+\mathcal{L}_{M}\right] \label{action BoxSP}.
\end{equation}
This action is singular according to the approach considered here. A similar action was analyzed
by Wands in \cite{Wands1994}. In his case, he introduced two scalar fields instead
of a scalar and a vector fields. If the field equations were analyzed in his
context, it would be possible to check that the two scalar fields would be dynamical
fields \cite{Felsager1983}. Here, the field equations indicate that only $\Phi$ and
$\phi_{0}$ are dynamical quantities while $\phi_{i}$ satisfy constraint equations.
So, in both cases there are just two additional degrees of freedom, showing the physical consistency
between Wands' approach and ours.

\subsection{Palatini formalism}

The starting point is
\[
S=\int d^{4}x\sqrt{-g}\left[  \mathcal{R}+\frac{c_{0}}{2}\mathcal{R}^{2}%
+\frac{c_{1}}{2}\nabla_{\mu}\mathcal{R}\nabla^{\mu}\mathcal{R}+\mathcal{L}%
_{M}\right]  .
\]
The tensorial fields are defined as above and the calculations lead to the
following expression for $S^{\prime}$:
\[
S^{\prime} = \int d^{4}x\sqrt{-g}\left[  \Phi\mathcal{R}-\frac{1}{2c_{0}}\left(
\Phi+\nabla_{\mu}\phi^{\mu}-1\right)  ^{2} \right.
- \left.\frac{\phi^{\mu}\phi_{\mu}}{2c_{1}%
}+\mathcal{L}_{M}\right]  .
\]
Substituting $\mathcal{R}$ in terms of $R$ leads to:
\[
S^{\prime} = \int d^{4}x\sqrt{-g}\left[  \Phi R+\frac{3}{2\Phi}\nabla_{\mu}%
\Phi\nabla^{\mu}\Phi \right.
- \left. \frac{1}{2c_{0}}\left(  \Phi+\nabla_{\mu}\phi^{\mu
}-1\right)  ^{2}-\frac{\phi^{\mu}\phi_{\mu}}{2c_{1}}+\mathcal{L}_{M}\right]  .
\]

The field equations are specified from (\ref{f SP}), (\ref{fprimesecond}) and
(\ref{eq mot 1})-(\ref{eq mot 3}) and lead to the following coupled equations for the scalar-vectorial part of $S^{\prime}$:
\[
\left\{
\begin{array}
[c]{l}%
\Phi=\frac{c_{0}}{\nabla_{\beta}\phi^{\beta}-1}\left[  \kappa T-\upsilon-\frac
{\phi^{\mu}\phi_{\mu}}{c_{1}}-\frac{1}{c_{0}}\left(  \nabla_{\beta}\phi
^{\beta}-1\right)  ^{2}\right] \, , \\
\frac{\phi_{\mu}}{c_{1}}-\nabla_{\mu}\left[  \frac{1}{c_{0}}\left(
\Phi+\nabla_{\gamma}\phi^{\gamma}-1\right)  \right]  =0 \, ,
\end{array}
\right.
\]
where $\upsilon$ is given by Eq. (\ref{vStarPod}). The first equation is a constraint equation for $\Phi$. Therefore, only $\phi_{0}$ satisfies a dynamical equation. This is different from what is obtained in the metric approach where both quantities are dynamical.

\subsection{Generalization: Starobinsky-Podolsky-higher-order action \label{sec-StaPodGen}}

The previous system may be generalized to:
\[
S = \int d^{4}x\sqrt{-g}\left[  R+\frac{c_{0}}{2}R^{2}+\frac{c_{1}}{2}%
\nabla_{\mu}R\nabla^{\mu}R\right.
+ \left. ...+\frac{c_{n}}{2}\nabla_{\mu_{1}}...\nabla
_{\mu_{n}}R\nabla^{\mu_{1}}...\nabla^{\mu_{n}}R+\mathcal{L}_{M}\right] \, ,
\]
which implies
\[
f\left(  \xi,\xi_{\mu},...\right)  =\xi+\frac{c_{0}}{2}\xi^{2}+\frac{c_{1}}%
{2}\xi_{\mu}\xi^{\mu}+...+\frac{c_{n}}{2}\xi^{\mu_{1}...\mu_{n}}\xi_{\mu
_{1}...\mu_{n}} \, .
\]
The condition
\[
\det H=\det\left(
\begin{array}
[c]{cccc}%
c_{0} & 0 & \cdots & 0\\
0 & c_{1}g^{\mu\nu} & \cdots & 0\\
\vdots & \vdots & \ddots & \vdots\\
0 & 0 & \cdots & c_{n}(g^{\mu\nu})^{n}%
\end{array}
\right)  \neq0
\]
is satisfied as long as $c_{0}\neq0,\,c_{1}\neq0,...,\,c_{n}\neq0$.

The results and conclusions are analogous to the ones in section \ref{Sec-Metric-Sta-Pod}, where the
potential takes the form:
\begin{eqnarray*}
U &  =& U\left(  \Phi,\phi_{\mu},...,\phi_{\mu_{1}...\mu_{n}},\nabla_{\mu}%
\phi^{\mu},...,\nabla_{\mu_{n}}...\nabla_{\mu_{1}}\phi^{\mu_{1}...\mu_{n}%
}\right)  \\
&  =& \frac{1}{2c_{0}}\left(  \Phi+\nabla_{\mu}\phi^{\mu}\right. \left. +...+\left(  -1\right)
^{n+1}\nabla_{\mu_{1}}...\nabla_{\mu_{n}}\phi^{\mu_{n}...\mu_{1}}-1\right)
^{2} +\frac{\phi^{\mu}\phi_{\mu}}{2c_{1}}+...+\frac{\phi^{\mu_{1}...\mu_{n}%
}\phi_{\mu_{1}...\mu_{n}}}{2c_{n}} \, .
\end{eqnarray*}

When considering the second order case, it was seen that by partial integration the higher order term $\nabla_{\mu}R\nabla^{\mu}R$ could be written as $R\Box R$.
Hence, one might wonder if it would be possible to proceed in a similar way for
the generalized case and obtain an action with only $R\Box^{n} R$ terms. The answer is no:
If the term $\nabla_{\mu_{1}}\nabla_{\mu_{2}}R\nabla^{\mu_{1}}\nabla^{\mu_{2}}R$ is considered,
by partial integration it is possible to verify that a Ricci tensor appears, i.e.
\begin{equation}
\nabla_{\mu_{1}}\nabla_{\mu_{2}}R\nabla^{\mu_{1}}\nabla^{\mu_{2}}%
R= R\square^{2}R+R\nabla^{\mu_{1}}\left(  R_{\mu_{2}\mu_{1}}\nabla^{\mu_{2}%
}R\right)+\partial_{\mu}S_{\left(  2\right)  }^{\mu}\label{WandRicci}
\end{equation}
For high order terms the situation is even more complicated because there appears Riemann tensors too. Thus, in the generalized case there is no equivalence between $f\left(  R,\nabla R,...,\nabla^{n}R\right)  $ and $f\left(  R,\Box R,...,\Box^{n}R\right)  $ theories.

\section{Final Remarks}

The equivalence of $f\left(  R,\nabla R,...\right)  $ theories and
scalar--multi-tensorial models has been studied in both metric and Palatini
formalisms. It has been demonstrated that, besides the
scalar field usually obtained in the equivalence of $f\left(  R\right)$ gravity to scalar-tensor theories, it is also necessary to introduce a tensorial field for each order of derivative of the scalar curvature. Moreover, it has been verified that when defining the scalar field as a
functional derivative, only the scalar field $\Phi$ is coupled to the scalar
curvature. The other tensor fields are minimally coupled to the gravitational field.

Both metric and Palatini approaches show that the scalar--multi-tensorial
theory from $f\left(  R,\nabla R,...\right)  $ gravity is a generalization of Brans-Dicke theory with $\omega_{0}=0$ or
$\omega_{0}=-\frac{3}{2}$, respectively. Here, beside the scalar field $\Phi$ introduced in Brans-Dicke theories, tensorial fields are also present in the potential $U$, being this a significant difference from regular Brans-Dicke approach. In the metric formalism, although no kinetic term for $\Phi$ is present in $S^{\prime}$,  this does not mean that no
dynamics is carried by $\Phi$. In the Palatini approach, the opposite
situation is found: even in the presence of the kinetic term, a constraint
equation is obtained for $\Phi$. These results are known on $f\left(
R\right)  $ gravity and are also valid for $f\left(  R,\nabla R,...\right)  $ theories.

It should be emphasized that $f\left(  R,\nabla R,...\right)$ theories are not the same as those coming from $f\left( R,\square R,...\right)$ Lagrangians \cite{Wands1994}. The former may differ from the last by terms
involving the Ricci and Riemann tensors, as shown, for instance, in Eq.(\ref{WandRicci}). Nevertheless, in particular cases $f\left(  R,\nabla R,...\right)$ gravity may reduce to $f\left(  R,\square R,...\right) $ models by taking appropriate contractions of indexes, e.g. $f\left(  R,\nabla ^{2} R\right)= g\left(  R,g^{\mu \nu} \nabla_{\mu} \nabla_{\nu} R\right) $. When this is the case, the resulting theory is likely to have a singular Hessian matrix so that the formalism developed in this work may not be directly applicable.

The field equations for $f(R)$ and $f(R,\nabla R,...)$ gravities
in the Brans-Dicke form\footnote{See Ref. \cite{Faraoni2010} for $f(R)$-gravity field
equations.} exhibit the same structure under the
generalization of $f^{\prime}$ to a functional derivative. Despite this similarity, almost all generic results of $f(R)$ gravity must be re-derived for specific applications such as  cosmology and the weak-field limit. An exception occurs with the Ehlers-Geren-Sachs (EGS) cosmology theorem \cite{Ehlers1968}. The EGS theorem states that \emph{if all observers see an isotropic radiation (like CMB) in the universe then the space-time is isotropic and spatially homogeneous and therefore it is described by the FLRW metric}.
As shown in Refs. \cite{Clarkson2001,Clarkson2003}, this theorem is valid for any scalar-tensor theory regardless the potential structure $U=U\left(  \Phi,\phi_{\mu},...,\nabla_{\mu}\phi^{\mu},...\right)$. Thus, for any $f(R,\nabla R,...)$ gravitational theory the description of a universe filled by an
isotropic CMB must be necessarily done with the FLRW line element. This fact was actually used in Ref. \cite{GRG2015} as an attempt to describe dark energy dynamics with a theory coming from an Einstein-Hilbert-Podolsky action of the type $f\left(  R,\nabla_{\mu}R\right)=R+\frac{c_{1}}{2}\nabla_{\mu}R\nabla^{\mu}R$.

A substantial difference between $f(R)$ and $f(R,\nabla R,...)$ actions concerns their
propagation modes. For example, besides the massless mode, the Starobinsky action
has only one massive mode of propagation corresponding to a positive square mass
\cite{Stelle1977}. On the other hand, Starobinsky-Podolsky action presents
positive (massive mode), negative (tachyon mode) and complex square masses
\cite{Modesto2016}. These features lead to important consequences such as instabilities
or lack of unitarity and, in principle, they could be used to constrain the physical actions.
These aspects are under consideration by the authors.

As a future work, it would be interesting to explore the consequences of $f(R,\nabla R,...)$ gravity to cosmology. Following a program similar to the one developed in the $f(R)$ context \cite{Faraoni2010}, one might study the general features of $f(R,\nabla R,...)$ cosmology in both metric and Palatini formalisms. This preliminary investigation would be an important step towards addressing more specific cosmological issues such as the present-day acceleration and the inflationary period.

\begin{acknowledgments}
The authors acknowledge FAPERN-Brazil for financial support. The authors thank two anonymous referees for their careful reading of the manuscript and useful comments that helped to improve the physical interpretation of the results presented in this work.
\end{acknowledgments}

\end{document}